\documentclass{article}
 \pdfoutput=1 
\usepackage{spconf,amsmath,graphicx}
                      
\usepackage{tikz}
\tikzstyle{pinstyle} = [pin edge={to-,thin,black}]
\usetikzlibrary{calc}
\newlength\fheight
\newlength\fwidth
\usepackage{pgfplots}
\usepackage{url}
\usepackage{array}
\newcolumntype{?}{!{\vrule width 1pt}}
\usepackage{booktabs}
\usepackage{stfloats}
\usepackage{multicol}
\usetikzlibrary{arrows,decorations.markings}
\usepackage{multirow}
\usepackage{arydshln}
\pgfplotsset{compat=newest}
\usepackage{amssymb}

\let\oldbibliography\thebibliography
\renewcommand{\thebibliography}[1]{%
  \oldbibliography{#1}%
  \setlength{\itemsep}{0pt}%
}

\title{Decoding Energy Modeling For Versatile Video Coding}
\name{Matthias Kr\"anzler, Christian Herglotz, and Andr\'e Kaup}
\address{Multimedia Communications and Signal Processing,\\
		Friedrich-Alexander University Erlangen-N\"urnberg (FAU)\\
		\{matthias.kraenzler, christian.herglotz, andre.kaup\}@fau.de\\
		Erlangen, Germany}

\usepackage{textcomp}
\usepackage[absolute]{textpos}

\newcommand{\copyrightstatement}{
    \begin{textblock}{15}(0.5,0.3)    
         \noindent
         \centering
         \textblockcolour{white}
         \footnotesize
         \copyright 2020 IEEE. Personal use of this material is permitted. Permission from IEEE must be obtained for all other uses, in any current or future media, including reprinting/republishing this material for advertising or promotional purposes, creating new collective works, for resale or redistribution to servers or lists, or reuse of any copyrighted component of this work in other works
    \end{textblock}
}

\begin{document}

\copyrightstatement

\maketitle

\begin{abstract}
In previous research, it was shown that the software decoding energy demand of High Efficiency Video Coding (HEVC) can be reduced by 15$\%$ by using a decoding-energy-rate-distortion optimization algorithm. To achieve this, the energy demand of the decoder has to be modeled by a bit stream feature-based model with sufficiently high accuracy. Therefore, we propose two bit stream feature-based models for the upcoming Versatile Video Coding (VVC) standard. The newly introduced models are compared with models from literature, which are used for HEVC. An evaluation of the proposed models reveals that the mean estimation error is similar to the results of the literature and yields an estimation error of 1.85$\%$ with 10-fold cross-validation.
\end{abstract}


\section{Introduction}
\label{sec:intro}

The constantly rising amount of internet traffic, especially of video data, shows the necessity to increase the bit rate efficiency of future video standards. According to~\cite{CSI2019}, the transmitted amount of data over the internet will increase by 26$\%$ each year. Simultaneously, the demand for video content with a higher resolution and a higher dynamic range is increasing. Therefore, the Joint Video Exploration Team (JVET) is working on a successor video standard to High Efficiency Video Coding (HEVC), which will be Versatile Video Coding (VVC). The first version of the VVC standard is expected to be finalized in 2020. The goal of VVC is the reduction of the bit rate in relation to HEVC by $50\%$ at equal subjective visual quality~\cite{Sullivan2017}.

The improvement of the rate-distortion (RD) efficiency is at the cost of higher computational complexity in terms of decoding energy demand, which is increased by over $80\%$ on average for the randomaccess (RA) configuration~\cite{JVET-Q0050}. The increased energy demand can be explained by new and computationally complex coding tools that have been introduced for VVC, e.g. decoder-side motion vector refinement (DMVR)~\cite{Gao2019}.

In previous research, it could be shown that the energy demand of an HEVC decoder can be reduced with a decoding-energy-rate-distortion optimization (DERDO) algorithm by $15\%$ at the cost of a bit rate increase by up to 5\%~\cite{HerglotzHeindelKaup}. For the optimization of the decoding energy demand, the energy has to be modeled with a bit stream feature-based model with sufficiently high accuracy. Corresponding feature models were developed, which can be used for DERDO for different video decoders such as H.264/AVC, VP9 \cite{HerglotzWenDaiEtAl2016}, and HEVC \cite{Kraenzler2019}.

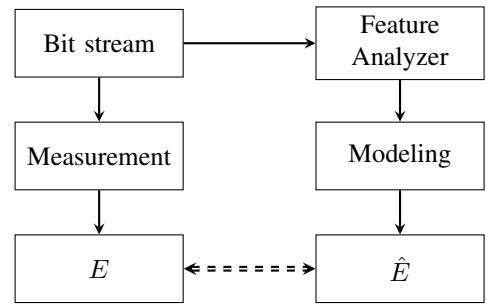
\begin{figure}[!t]

	\centering
	\setlength\fheight{4cm}
	\setlength\fwidth{7.5cm}
	\begin{tikzpicture}
	\begin{scope}[minimum width=15mm,minimum height=9mm,align=center]
	\node[draw,text width=2.0cm] (v1) at (0,0) {Bit stream};
	\node[draw,text width=2.0cm] (v2) at ($(v1)+(4,0)$) {Feature Analyzer};
    \node[draw,text width=2.0cm] (v3) at ($(v1)+(0,-1.5)$) {Measurement};
	\node[draw,text width=2.0cm] (v4) at ($(v3)+(4,0)$) {Modeling};
	\node[draw,text width=2.0cm] (v5) at ($(v3)+(0,-1.5)$) {$E$};
	\node[draw,text width=2.0cm] (v6) at ($(v5)+(4,0)$) {$\hat{E}$};

	\end{scope}

	\draw[-stealth,thick] (v1) -- (v2);
	\draw[-stealth,thick] (v1) -- (v3);
	\draw[-stealth,thick] (v3) -- (v5);
	\draw[-stealth,thick] (v2) -- (v4);
	\draw[-stealth,thick] (v4) -- (v6);

	\draw[stealth-stealth,double,thick,dashed] (v5) -- (v6);
    
	\end{tikzpicture}
	   \vspace{-0.25cm}
	\caption{Overview of the decoding energy modeling evaluation setup for VVC coded bit streams.}
	\label{fig:BlockChart}
		   \vspace{-0.5cm}
\end{figure}

In this paper, two new models are proposed to model the energy demand of VVC with an estimation error of less than 5\%, which is considered to be sufficiently accurate to enable DERDO. The different parts of the evaluation setup for the energy modeling are shown in Figure~\ref{fig:BlockChart}. First, bit streams are encoded with the reference software for VVC. Afterwards, the decoding energy $E$ of the bit streams is measured. Then, the bit streams are analyzed with a bit stream feature analyzer (BSFA) software \cite{VTMAnalyzer}, which extracts all features that are needed to model the estimated decoding energy demand $\hat{E}$. The goal of the modeling is that $\hat{E}$ and $E$ are equal.

This paper is organized as follows. Section~\ref{sec:measurement} explains the performed measurements and the test set of bit streams for the evaluation of the estimation accuracy. Afterwards, we give a brief overview of bit stream feature based models. Then, two new models for VVC will be introduced. The results of the decoding energy modeling for HEVC and VVC will be discussed in Section~\ref{sec:evaluation}. Finally, Section~\ref{sec:conclusion} concludes the paper.

\section{Measurement and Test Set}
\label{sec:measurement}

In this section, we will explain at first the measurement setup and then the used test sequences. For the measurement of the decoding energy demand, we perform two measurements. The first energy measurement determines the demand during the decoding process. Afterwards, the second measurement determines the energy demand during the idle mode over the same duration as the decoding process. With these two consecutive measurements, it is possible to determine the energy that corresponds to the decoding energy demand by calculating the difference between both measurements.

For energy measurements, we use a desktop PC with an Intel i7-8700 CPU. The Intel CPU has an integrated power meter called Running Average Power Limit (RAPL). This power meter can measure the power demand of the CPU socket. Our measurements with RAPL are based on the explanations of \cite{HerglotzSpringerReichenbachEtAl2018}. We can observe that the measurements are influenced, e.g. by background processes of the operating system. Furthermore, due to noise, we assume that one measurement is not sufficient for the determination of the energy. Therefore, we use the statistical test described in \cite{HerglotzSpringerReichenbachEtAl2018} to evaluate the statistical correctness of the measurements with a confidence interval test. This test is defined by
\begin{equation}
2 \cdot \frac{\sigma}{\sqrt{m}} \cdot t_{\alpha} \cdot \left( m - 1 \right)
< \beta \cdot \overline{E},
\label{eq:confidence}
\end{equation}
where $\sigma$ is the standard deviation of the measurement, $m$ the number of measurements, which has a minimum value of five, $\beta$ the allowed deviation of the actual energy, $t_{\alpha}$ the critical t-value of the Student's t-distribution, and $\alpha$ the probability that the mean measured energy $\overline{E}$ is within the allowed deviation. In this paper, the value of $\alpha$ is 0.99 and of $\beta$ is 0.02, which means that the actual energy corresponds to $\overline{E}$ with a certainty of 99$\%$ within a deviation of less than 2$\%$.

\begin{table*}[!t]
\caption{List of all tools that are switched off to generate additional bit streams for the modeling. For each tool, the corresponding configuration and number of bit streams are given. A detailed description for each tool can be found in the references.}
\label{tab:ToolOff}
\begin{center}
\vspace{-0.2cm}
\begin{tabular}{l | c | c | c | c}
Switched off Tool  & Acronym & Reference & Configuration & \#Bit Streams  \\
\hline \hline
Adaptive Loop Filter                  & ALF   & \cite{Tsai2013}            & RA & 92 \\
Bi-directional optical flow           & BDOF  & \cite{Alshin2010}          & RA & 92 \\
Decoder-side motion vector refinement & DMVR  & \cite{Gao2019}             & RA & 92 \\
Intra Sub-Partitions                  & ISP   & \cite{Luxan-Hernandez2019} & AI & 104 \\
Low frequency non-separable transform & LFNST & \cite{Koo2019}             & RA & 92 \\
Matrix weighted intra prediction      & MIP   & \cite{Schaefer2019}        & AI & 104 \\
Multiple Transform Selection          & MTS   & \cite{Nguyen2019}          & RA & 92 \\
Triangular Partition Mode             & TPM   & \cite{Blaeser2019}         & RA & 92 \\
\end{tabular}
\end{center}
\vspace{-0.8cm}
\end{table*}

For the decoding of the bit streams, we use VTM-7.0 \cite{VTM7}, which is the reference software implementation of VVC. For comparison of the modeling accuracy, we use the reference software of HEVC, which is HM-16.20~\cite{HM1620}. For both software implementations, we code the bit streams according to the common test conditions (CTC) for standard dynamic range (SDR) content~\cite{JVET-N1010}. 

Furthermore, we encode additional bit streams by switching off several coding tools according to the explanations of \cite{JVET-Q0013}. Like that, the size of the training set can be increased, which reduces overfitting due to a high number of parameters. In Table~\ref{tab:ToolOff}, considered tools are switched off. A detailed explanation of each coding tool can be found with the corresponding source in Table~\ref{tab:ToolOff}. In total, we have encoded 760 bit streams with a tool switched off.

\section{Bit Stream Feature Based Modeling}
\label{sec:features}
In this section, at first bit stream feature-based modeling will be explained in general. Afterwards, we will introduce the proposed feature-based versatile (FV) and the feature-based versatile simple (FVS) model for VVC.

For the estimation, we use bit-stream feature based modeling, which is defined by
\begin{equation}
\hat{E} = \sum_{\forall j} e_j \cdot n_j ,
\end{equation}
where $j$ corresponds to a bit stream feature, $e_j$ to the specific energy coefficient of the feature $j$, and $n_j$ to the number of occurrences of the feature within the bit stream. An example for a feature is the number of residual coefficients. For this feature, $e_{\text{coeff}}$ corresponds to the energy that is required to parse one nonzero coefficient and $n_{\text{coeff}}$ to the number of nonzero coefficients in the bit stream.

In order to model $\hat{E}$, we have to derive $n_j$ for each feature and bit stream in the test set. Therefore, we use the BSFA software for VVC, which is online available \cite{VTMAnalyzer}. Then, we measure $E$ for each bit stream with a series of measurements. Finally, we train $e_{j}$ by using a trust-region-reflective algorithm with least-squares curve fitting \cite{ColemanLi1996}.

All features of the bit stream models can be assigned to different categories of a hybrid video coder (e.g. transformation). A detailed description of each category can be found in \cite{Kraenzler2019}. Furthermore, all features are counted in different levels, which provide information about the part of the video that is counted. We distinguish between the levels Slice, Pel, Coding Tree Blocks (CTB), Boundary, and Blockpel. For Slice, we count the number of the corresponding slice type, for Pel the number of pixels, and for CTBs the number of blocks, respectively. The feature Val is a special case of the Pel category, which adds up the value of all coefficients with the binary logarithm. For the feature BS0, we count the boundaries that are filtered by a deblocking filter in the level Boundary. 

In VVC, the multi-type tree partitioning allows different types of rectangular block shapes. Therefore, the BSFA software counts the block based features of VVC (e.g. IntraBlocks) according to the following matrix:
\begin{equation}
\begin{small}
\left( \begin{array}{c c c c c}
			1 \times 1   & 2 \times 1   & \cdots  & 64 \times 1   & 128 \times 1 \\
			1 \times 2   & 2 \times 2   & \cdots  & 64 \times 2   & 128 \times 2 \\
			\vdots       & \vdots       & \ddots  & \vdots        & \vdots \\
			1 \times 64  & 2 \times 64  & \cdots  & 64 \times 64  & 128 \times 64 \\
			1 \times 128 & 2 \times 128 & \cdots  & 64 \times 128 & 128 \times 128 \\
\end{array}\right) ,
\end{small}
\end{equation}
where the first dimension corresponds to the width of the block and the second dimension to the height of the block, respectively. Each dimension is defined between 1 and 128 pels in integer powers of 2. For a specific feature (e.g. MIP), the frequency of occurrences is counted of corresponding block size in the matrix. 

\begin{table}[!t]
\caption{List of all features of the FV and FVS model. For each feature, the index of the feature and the corresponding counting level is given. Furthermore, the features are grouped according categories.}
\label{tab:List}
\vspace{-0.3cm}
\begin{center}
\begin{tabular}{ c |  l  | c | c | c}
	Index $j$ &	Feature Name	       & Level   & FV & FVS\\
\hline 
	\multicolumn{5}{c}{General} \\
	\hline \hline 
 1	    & EO	               & -     & $\checkmark$ & $\checkmark$\\
 2	    & ISlice	           & Slice & $\checkmark$ & $\checkmark$\\
 3	    & PSlice	           & Slice & $\checkmark$ & -\\
 4	    & BSlice	           & Slice & $\checkmark$ & -\\
 -      & PBSlice              & Slice & -            & $\checkmark$\\
\hline	
	      \multicolumn{5}{c}{Intra} \\
	      \hline \hline 
 ~~5-17    & IntraBlocks	       & Blockpel & $\checkmark$ & $\checkmark$\\
 18-30	& ISP                    & Blockpel & $\checkmark$ & -\\ 
 31-43	& IntraPDPC	           & Blockpel & $\checkmark$ & - \\
 44-56	& MIP        	           & Blockpel & $\checkmark$ & -\\
 57-69	& IBC	           & Blockpel & $\checkmark$ & -\\
 \hline
	      \multicolumn{5}{c}{Inter} \\
	      \hline \hline 
 70-82	    & InterInter	       & Blockpel & $\checkmark$ & -\\
 83-95	& InterMerge	       & Blockpel & $\checkmark$ & -\\
 -       	& InterCU	           & Blockpel & - & $\checkmark$\\
 ~~96-108     & InterSkip	           & Blockpel & $\checkmark$ & $\checkmark$\\
 109-121	& Affine	           & Blockpel & $\checkmark$ & -\\
 122-134	& TriangleSplit	       & Blockpel & $\checkmark$ & -\\
 135-147    & DMVR                 & Blockpel & $\checkmark$ & -\\
 148-160    & BDOF                 & Blockpel & $\checkmark$ & -\\
 161	    & Uni	               & Pel & $\checkmark$ & $\checkmark$\\
 162	    & Bi 	               & Pel & $\checkmark$ & $\checkmark$\\
 163	    & FracPelHor	       & Pel & $\checkmark$ & $\checkmark$\\
 164	    & FracPelVer	       & Pel & $\checkmark$ & $\checkmark$\\
 165	    & FracPelBoth	       & Pel & $\checkmark$ & $\checkmark$\\
 166	    & CopyPel	           & Pel & $\checkmark$ & $\checkmark$\\
 \hline
	      \multicolumn{5}{c}{Transform} \\
	      \hline \hline 
 167-179	& Transform	           & Blockpel & $\checkmark$ & $\checkmark$\\
 180-192	& TransformSkip        & Blockpel & $\checkmark$ & -\\
 193-205	& TransformNoCbf       & Blockpel & $\checkmark$ & -\\
 206-218	& LFNST	               & Blockpel & $\checkmark$ & -\\
 219	    & Coeff	               & Pel & $\checkmark$ & $\checkmark$\\
 220	    & CoeffG1	           & Pel & $\checkmark$ & - \\
 221	    & Val	               & Pel (Log) & $\checkmark$ & $\checkmark$\\
 \hline 
	      \multicolumn{5}{c}{In-Loop Filter} \\
	      \hline \hline 
 222-224	    & BS0~-~2	               & Boundary & $\checkmark$ & -\\
  -	        & BS	               & Boundary & - & $\checkmark$\\
 225	    & SAO Luma BO	           & CTB & $\checkmark$ & -\\
 226	    & SAO Luma EO	           & CTB & $\checkmark$ & -\\
 227	    & SAO Chroma BO	           & CTB & $\checkmark$ & -\\
 228	    & SAO Chroma EO	           & CTB & $\checkmark$ & -\\
  -  	    & SAO	               & CTB & - & $\checkmark$\\
 229	    & ALF Luma              & CTB & $\checkmark$ & -\\
 230	    & ALF Chroma	               & CTB & $\checkmark$ & -\\
  -  	    & ALF	               & CTB & - & $\checkmark$\\
\end{tabular}
\end{center}
\vspace{-1cm}
\end{table}

However, this matrix of block sizes is not practical for the modeling of the decoding energy, because each feature would have 64 levels resulting in significant overfitting. Therefore, the block sizes are transformed to the following vector with 13 entries:
\begin{small}
\begin{center} $\left( \begin{array}{ccccccc}
   4 & 8 & 16 & \hdots & 4096 & 8192 & 16384 \\
\end{array}                    \right)$, 
\end{center}
\end{small}
where each entry corresponds to the number of pels within the block. All blockpel-level features are counted with respect to these different block sizes.

In Table~\ref{tab:List}, all features of the FV and FVS model are listed. For each feature, the index $j$ for the FV model, the corresponding category and the counting level are given. In total, the FV model has 230 features and the FVS model has 67 features. The proposed models are extensions of the feature-based simple (FS) \cite{HerglotzSpringerReichenbachEtAl2018} and the feature-based universal (FU) \cite{Kraenzler2019} model, which were developed for HEVC. The FV model is an extension of FU, and FVS of FV. In literature, the FS model is used in DERDO and the FU model is currently the most accurate model available for HEVC. The FS model has 27 features and the FU model 100 features. 

In comparison to the FU model, the following features are newly introduced. With the feature IntraBlocks, we count all blocks that are intra predicted. Furthermore, the features ISP and MIP count the blocks, where the tools ISP and MIP are used. With IntraPDPC, all blocks are counted if position dependent prediction combination (PDPC) is applied and for IBC, if the tool intra block copy (IBC) is used. A detailed description of the last two tools can be found in \cite{JVET-P2002}. 

For the features DMVR, BDOF, LFNST, and ALF, we count the corresponding coding tools of Table~\ref{tab:ToolOff}. The feature Affine counts all blocks that use affine motion compensation. The coding tool TPM is counted by the feature TriangleSplit and TransformNoCBF comprises all transform blocks without nonzero coefficients. The in-loop filter features BS counts all boundaries that are filtered by a deblocking filter and SAO counts CTBs that are filtered by the sample adaptive offset (SAO) filter.

\vspace*{-.2cm}
\section{Evaluation}
\label{sec:evaluation}
\vspace{-.2cm}
\begin{table}[!t]
\caption{Evaluation of the different bit stream feature-based models for HEVC and VVC. For HEVC, we use the FS and the FU model, and for VVC we use the FVS und FV model.}

\label{tab:eval10fold}
\def\arraystretch{1}
\vspace{-0.3cm}
\begin{center}
\begin{tabular}{c | c |  c |  c |  c |  c}
 & & \multicolumn{2}{c|}{HEVC} & \multicolumn{2}{c}{VVC} \\
 Decoder & Test Set      & FS & FU & FVS &FV  \\
\hline \hline
HM                      & CTC    & $2.73\%$  & $1.76\%$& - & -   \\
\hdashline
\multirow{2}{*}{VTM}    & CTC    & - & -   & $3.95\%$  & $2.63\%$ \\
                        & Merge  & - & -   & $3.04\%$  & $1.85\%$ \\
\end{tabular}
\end{center}
\vspace{-0.8cm}
\end{table}

For the evaluation of the modeling, we use the following equation to determine the mean relative estimation error 

\begin{equation}
\overline{\varepsilon} = \frac{1}{N} \cdot \sum_{n=1}^{N} \left|  \frac{\hat{E}_n - E_n}{E_n} \right| ,
\end{equation}
where $N$ corresponds to the number of bit streams within the setup, $\hat{E}_n$ is the estimated energy of the modeling for the $n$-th bit stream, and $E_n$ is the measured energy. For the training of $e_j$, we use the 10-fold cross-validation. Therefore, the bit streams of the setups are randomly separated into \mbox{10-subsets}. For each subset, 9 subsets are used for training and the remaining subset is used for validation. 

The evaluation is performed with different sets of bit streams. The setup CTC includes all bit streams that are encoded according to the CTCs mentioned in Section~\ref{sec:measurement} and has 276 bit streams. For VVC, we merge the bit streams of the CTC setup with the 760 bit streams that have a tool switched off. This test set is called Merge. 

In Table~\ref{tab:eval10fold}, the results for all models are given. It is observed that the estimation error is lower than $5\%$ for all models. For HEVC, the FS model has a $\bar{\varepsilon}$ of 2.73$\%$ and the FU model a $\bar{\varepsilon}$ of 1.76$\%$, respectively. For the modeling of VVC with the CTC setup, $\bar{\varepsilon}$ is $3.95\%$ for the FVS model and $2.63\%$ for the FV model. Therefore, the estimation error of the FVS model for VVC is increased significantly in comparison to the modeling of the FS model for HEVC. Considering the Merge setup, the estimation error $\bar{\varepsilon}$ is reduced to $3.04\%$ for the FVS model. Therefore, the modeling of the FVS model has a higher estimation error $\bar{\varepsilon}$ than the FS model. This observation can be explained by the fact that VVC has coding tools that are not represented by the FVS model, e.g. DMVR or BDOF. 

For the Merge setup, the estimation error $\bar{\varepsilon}$ of the FV model is $1.85\%$. The decreased value of the estimation error $\bar{\varepsilon}$ for the Merge set compared to the CTC set indicates that the number of bit streams have significant influence on the modeling accuracy, which is due to a relatively small training set (276) in comparison with a large number of parameters (230). Comparing the results of the FU model for HEVC and the FV model for VVC shows that both models have a similar estimation accuracy. Furthermore, we observe that the estimation error $\bar{\varepsilon}$ is lower than the allowed deviation ($\beta = 2\%$) of the statistical test. 

\begin{figure}[!t]
\input{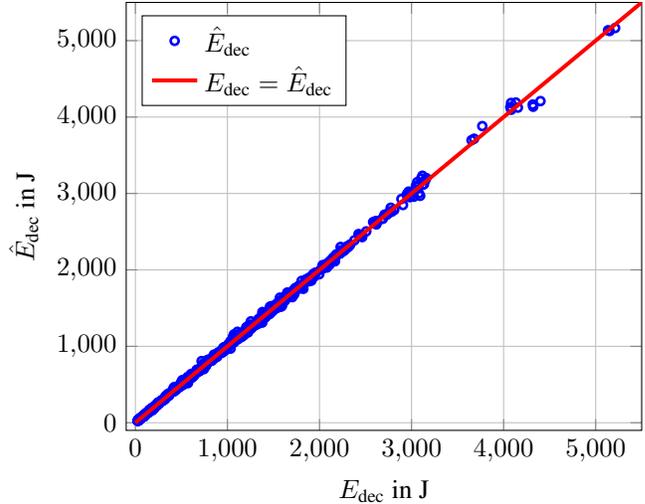}
\vspace{-0.8cm}
\caption{Evaluation of the FV model for the combined setup. Each bit stream of the setup corresponds to a blue marker. The red line corresponds to the aim of the modeling, where $E$ is equal to $\hat{E}$.}
\label{fig:modeling}
\vspace{-.6cm}
\end{figure}

Figure~\ref{fig:modeling} compares the estimated with the measured energies of the FV model  in a scatter plot. Each blue marker in the figure corresponds to one bit stream of the setup. On the x-axis, the measured decoding energy $E$ is given in Joule and on the y-axis, the estimated decoding energy $\hat{E}$ is given in Joule, respectively. The red line shows the ideal modeling of the energy, where $\hat{E}$ is equal to $E$. From the graph it can be determined that the estimation of the bit streams has a high modeling accuracy.

\vspace{-0.2cm}
\section{Conclusion}
\label{sec:conclusion}

In this paper, we proposed an accurate model with more than 200 features and a simple model with 67 features. The accurate model comprises new features that represent new coding tools that are introduced into VVC. Although the estimation error of the simple model is larger than the estimation error of the reference model for VVC, which can be explained by missing coding tools, mean error below 4$\%$ can still be achieved. However, to obtain more accurate estimates, a more complex model was proposed with an estimation error of 1.85$\%$, which is similar to the results for HEVC.

In future work, the modeling of VVC can be used to optimize the energy demand of bit streams that are decoded by a VVC software decoder. Therefore, the DERDO algorithm that was proposed in~\cite{HerglotzHeindelKaup} for HEVC can be adapted to VVC. Furthermore, the model will be validated on consumer decoder implementations.

\clearpage
\newpage

\bibliographystyle{IEEEbib}

\end{document}